\documentclass[aps,prb,twocolumn,showkeys,showpacs]{revtex4-1}

\usepackage{amsmath}
\usepackage{amsfonts}
\usepackage{amssymb}
\usepackage{feyn}
\usepackage{graphicx}

\newcommand{\ee}{\mathbf{e}}
\newcommand{\ff}{\mathbf{f}}
\newcommand{\jj}{\mathbf{j}}
\newcommand{\kk}{\mathbf{k}}
\newcommand{\qq}{\mathbf{q}}
\newcommand{\rr}{\mathbf{r}}
\newcommand{\vv}{\mathbf{v}}
\newcommand{\pp}{\mathbf{p}}

\newcommand{\XX}{\mathbf{X}}
\newcommand{\RR}{\mathbf{R}}
\renewcommand{\AA}{\mathbf{A}}
\newcommand{\BB}{\mathbf{B}}
\newcommand{\EE}{\mathbf{E}}
\newcommand{\QQ}{\mathbf{Q}}
\newcommand{\PP}{\mathbf{P}}

\newcommand{\RRR}{\boldsymbol{\mathcal{R}}}

\newcommand{\nblk}{\mathbf{\nabla_k}}
\newcommand{\nblr}{\mathbf{\nabla_r}}
\newcommand{\nblR}{\mathbf{\nabla_R}}

\newcommand{\sign}{\mathrm{sign}}
\renewcommand{\Re}{\mathrm{Re}}
\renewcommand{\Im}{\mathrm{Im}}

\begin{document}

\title{Quantum Kinetics of the Magneto--Photo--Galvanic Effect}

\author{Dieter \surname{Hornung}$^1$ and Ralph \surname{von~Baltz}$^{2}$}
\email[Corresponding author:\;]{ralph.baltz@kit.edu}
\affiliation{$^1$Department of Mechatronics, Faculty of Engineering, 
                 University of Applied Sciences, 66117 Saarbr\"ucken, Germany}
\affiliation{$^2$Institute for Theory of Condensed Matter, Faculty of Physics,
                 Karlsruhe Institute of Technology (KIT), 76131 Karlsruhe, Germany}

\date{Mai 27, 2021}

\begin{abstract}
Using the Keldysh technique, we derive a set of quasiclassical equations for Bloch electrons in noncentrosymmetric
crystals upon excitation with quasimonochromatic radiation in the presence of external electrical and magnetic fields.
These equations are the analog to the semiconductor--Bloch--equations for the dynamics of electrons including
the photogalvanic effect (PGE) in particular the shift mechanism.
The shift PGE was recently identified as showing promise for the development of new photovoltaic materials.
In addition, our theory may be useful to investigate the interplay between breaking time--reversal symmetry and
topological properties as well as the analysis of recent local excitation experiments in nanophotonics.
Explicit results for the photogalvanic tensors are presented for linear and circular polarized light and a magnetic field.
In addition, we disprove existing statements that the shift--photogalvanic effect does not contribute to the
photo--Hall current.
\end{abstract}



\maketitle

\section{Introduction}
In noncentrosymmetric crystals a direct current can be induced upon the absorption of light under homogeneous
conditions. This phenomenon was discovered more than 50 years ago and it was termed  the bulk photovoltaic
effect (BPVE) or the photogalvanic effect (PGE), cf. Sturman and Fridkin\cite{Sturman-1}.
As a result of two major discoveries the PGE recently gained an unprecedent boost: the discovery of ferroelectric perovskite
materials\cite{Akihiro} in 2009 as potentially relevant solar cell materials and the discovery of Weyl semimetals in
2015 with topologically protected states\cite{Xu}. The underlying physics is intimately connected with the so--called
shift mechanism (as described later).
The aim of this paper is to work out a semiclassical theory for the PGE
which is suited for numerical investigations including external electrical and magnetic fields.

The PGE depends on the properties of the material, applied fields and the properties of the absorbed light.
At first order in the light intensity and in an external magnetic field with induction $\BB$,
symmetry requires the following representation for the radiation--induced direct current
(no static electrical field, neglecting photon momentum):
\begin{eqnarray}
  j_\alpha &=& I\; \big( P^S_{\alpha\mu\nu}(\omega) + R^S_{\alpha\beta\mu\nu}(\omega)\; B_\beta \big)\; \mathrm{Re}(e^*_{\mu}
   e_\nu)\nonumber\\
   &&+ I\; \big( P^A_{\alpha\mu\nu}(\omega) + R^A_{\alpha\beta\mu\nu}(\omega)\; B_\beta \big)\; \mathrm{Im}(e^*_{\mu}e_\nu).
 \label{j}
\end{eqnarray}
Symbols have the following meaning: $I$ (local) intensity, $\omega$ frequency, $e_{\mu}$ (Cartesian) components of the
(complex) unit polarization vector $\ee$ of the light.
Indices $\alpha,\beta,\mu,\nu \in \left \{ x,y,z \right \}$ indicate cartesian components; an asterisk indicates complex conjugation.
$\PP^S$ and $\PP^A$ denote polar tensors of rank three whereas $\RR^S$ and $\RR^A$ are of rank four with axial symmetry.
Superscripts S and A specify symmetry and antisymmetry with respect to polarization indices $\mu,\nu$, and their contributions
are usually termed ``linear'' and ``circular'', respectively\cite{comment-1}.
$\PP^S$ is analogous to the piezotensor whereas $\PP^A$ is equivalent to the (rank two axial) gyrotensor in gyrotropic
media, and $\RR^A$ is equivalent to a polar tensor of rank three, see Birss\cite{Birss}.

In the spirit of nonlinear optics\cite{Boyd}, the  photogalvanic (PG) current results from a quadratic term in the current--field relation.
Standard second--order quantum mechanical response theory\cite{RvB-2} revealed two different origins of the PGE:
a ``ballistic'' (kinetic) mechanism and a ``shift'' mechanism. The ballistic PGE results from asymmetric optical transitions
in cooperation with impurities or phonon scattering, which is described by the diagonal matrix elements of the density
operator (with respect to a Bloch basis).
The shift PGE, on the other hand, is a band structure property and results from the nondiagonal elements.
It is intimately related to the Bloch representation of the position operator\cite{Blount}, which leads
to a shift of Bloch wave packets in real space upon optical transitions\cite{RvB-3,Belinicher-2,Kristoffel-1}.
The circular PGE ($\PP^A$ term) is invariant under time reversal as opposed to the linear PGE ($\PP^S$ term),
in which an external magnetic field breaks time reversal explicitly.

For linear polarized light, the shift--current contribution can be represented as
($\PP^S$ term, a reformulation of Eq.~(19) of Ref.\cite{RvB-3})
\begin{eqnarray}
 \mathbf j_{\mbox{\tiny PG}} &=& \frac{I}{\hbar\omega} \frac{e^3}{4\pi^2\,\omega\,m_0^2\,\epsilon_0\,c\,\eta}
 \int (f_{v,0}-f_{c,0})\nonumber\\
&&\times\left| \langle c,\kk \left|\ee\cdot\pp \right| v,\kk \rangle\right|^2\,\;\mathbf{s}_{cv}(\ee,\kk)\nonumber\\
&&\times\delta ( E_c(\kk)-E_v(\kk)-\hbar\omega )\; d^3k ,\label{shiftcur}\\
\mathbf{s}_{cv}(\ee,\kk) &=& {\mathbf X}_{vv}(\kk) - \mathbf{X}_{cc}(\kk) +\nblk\Phi_{cv}(\ee,\kk),\label{shiftvec}\\
\mathbf{X}_{mn}(\kk)&=& \int{i\; u^*_{m\kk}(\rr)\mathbf{\nabla_k} u_{n\kk}(\rr)\, d^3r},\label{Berry-X}
\end{eqnarray}
where $\Phi_{cv}(\ee,\kk)$ is defined via the expression
\begin{equation}
 \langle c,\kk \left|\ee\cdot\pp \right| v,\kk\rangle = i\,\left| \langle c,\kk \left|\ee\cdot\pp \right| v,\kk \rangle\right|
 \, e^{i\,\Phi_{cv}(\ee,\kk) }.\label{p-matrix}
\end{equation}
$|n,\kk\rangle$ denotes the Bloch states of (conduction and valence) bands\cite{twobands} $n=c,v$ at wave vector $\kk$,
$E_n(\kk)$ is the band energy, and $u_{n\kk}(\rr) = (\rr|n,\kk)$ is the lattice--periodic part of the Bloch function
$\langle \rr|n,\kk\rangle$.
$f_{n,0}(\kk)$ is the equilibrium Fermi function, $m_0$ is the free--electron mass and $e$ is the elementary charge.
$\ee$ (real) denotes the polarization vector, and $I$ is the local intensity of the radiation at frequency $\omega$.
$\eta$ is the refractive index of the material, and integrals over $r$ and $k$ extend over the crystal unit cell and the
Brillouin zone, respectively. Note that the shift current does not depend on the carrier mobility.

By construction, the shift vector $ {\bf s}_{cv}(\ee,\kk)$ is invariant with respect to phase transformations of the
Bloch states, however, it depends on the polarization of the light, and therefore, it is not a genuine property of
 the material (in contrast to $\PP^S$).
Second--order quantum response theory was fully exploited by Sipe and collaborators\cite{Sipe-1}, who developed
a nowadays widely used approach to study nonlinear optical phenomena on a microscopic level, such as 
second--order--harmonic generation and the shift PGE.
Results (\ref{shiftvec})--(\ref{p-matrix}) are valid for only linear polarization and they are implicitely contained
in Ref.\cite{Sipe-1}(Eq.~(58) and below, linear polarization of arbitrary direction).
The shift distance is comparable to the crystal unit cell\cite{Sipe-3,Hornung-1} and may be even larger,
e.g. CdSe: 0.4 nm, GaP: 0.9 nm.

Up to 2006 (to the best of our knowledge) there was only one band structure evaluation\cite{Hornung-1} of
Eq.~(\ref{shiftcur}) which was performed for n--doped GaP.
This material has been used as a fast and robust IR monitor\cite{Gibson}.
First principles band structure calculations were performed by Nastos and Sipe\cite{Sipe-2,Sipe-3}
for GaAs and GaP below and above the band gap and for CdSe and CdS.
Young and Rappe\cite{Rappe-1} confirmed the shift mechanism as given by Eqs.~(\ref{shiftcur}--\ref{Berry-X}) for some
``old materials" like BaTiO$_3$ and KNbO$_3$  and claimed its key role in the high efficiency of the
new ferroelectrics in solar energy conversion of up to 23\%, see e.g. Refs.\cite{Young,Wei,Paillard,Lopez,Liang,Cook,Ogawa}.
Recent numerical studies have discovered new groups of promising materials with large
shift contributions up to 20 times higher than previously known\cite{typical}, 
e.g. the quasi--two--dimensional systems GeS\cite{Rangel} and MoS$_2$\cite{Schankler},
chiral materials\cite{Zhang-b}, and materials using strain engineering\cite{Zhao}.

It became obvious that the shift vector equation~(\ref{shiftvec}) is a Berry connection which provides a sensitive
tool to analyze the topological nature of quantum states in the recently discovered Weyl semimetals (see, e.g.,
\hbox{Refs.\cite{Koenig,Zhang,Ma,Parker}}).
A recent revisit of the second--order optical response by Holder et al.\cite{Yan} identified three different
mechanisms to generate a dc current:
the Berry curvature, a term closely related to the quantum metric, and the diabatic motion.
Berry connections have also been recognized as relevant ingredients for the quasiclassical dynamics of
Bloch electrons\cite{Sundaram} and the anomalous Hall effect\cite{Sinitsyn}.
Other interesting phenomena and applications with relation to the shift mechanism are, e.g.,
(i) FIR detectors in the form of semiconductor heterostructures\cite{Schneider},
(ii) the shift vector as the geometrical origin of beam shifts\cite{Li-kun},
(iii) nanotubes\cite{Kral}, and
(iv) twisted graphene bilayers\cite{Gao}.

By using the Keldysh technique we derive a set of quasiclassical equations for the PGE (Sect.~II) upon (inhomogeneous)
excitation and including external electrical and magnetic fields.
Our theory relies on the following assumptions:
(i) electron Bloch-states are a relevant basis,
(ii) scattering and recombination are treated on a phenomenological level, and
(iii) electron--hole Coulomb--interaction is neglected.
Explicit results for the PG tensors are worked out in Sec.~III.
Section~IV gives a summary and discussion whereas, Appendixes A--C contain technical details and an application to GaP.

\section{Quantum Kinetics}
The quantum kinetic theory of the PGE is based on a Hermitian matrix function $\ff$ with elements $f_{mn}(\kk,\RR,T)$
which describes the single--particle states of the crystal, $m$ and $n$ denote band indices. The arguments of $\ff$ are,
besides the wave vector $\kk$, the position vector $\RR$ and the time $T$. This theory is a generalization of the classical
Boltzmann description; it includes, however, diagonal (local electron concentrations) as well as nondiagonal
(nondissipative, coherent) contributions of the density operator.

The basic equations for $\ff$ are derived by using the Keldysh technique as formulated by Rammer and Smith\cite{Rammer}.
This technique provides a consistent way to construct a quasiclassical description at finite temperatures; it uses solely
gauge invariant quantities. External fields can easily be included, and applications are much simpler to work out than
a full quantum mechanical treatment as in Eqs.~(\ref{shiftcur}--\ref{Berry-X}).

\subsection{Keldysh formulation}

It is algebraically favorable to use a representation in which all Keldysh matrices have the Jordan normal form
(Ref.~\cite{Rammer}, Sec.IIB). For example the Green's function $\hat{\mathbf G}$ reads
\[
\hat{\mathbf G} = \left[
\begin{tabular}{ll}
$G^R$ & $G^K$\\
$0$   & $G^A$
\end{tabular}
\right].
\]
$G^R$ and $G^A$ denote the usual retarded and advanced Green's functions and $G^K$ is the Keldysh function, which plays
a crucial role in this formulation,
\begin{eqnarray*}
G^R(\RR,T;\rr,t) &=& +\theta(t) \{ G^>(\RR,T;\rr,t) - G^<(\RR,T;\rr,t)\},\\
G^A(\RR,T;\rr,t) &=& -\theta(-t)\{ G^>(\RR,T;\rr,t) - G^<(\RR,T;\rr,t)\},\\
G^K(\RR,T;\rr,t) &=& G^>(\RR,T;\rr,t) + G^<(\RR,T;\rr,t).
\end{eqnarray*}
All these functions are special combinations of the Kadanoff--Baym functions $G^<$ and $G^>$
(Ref.~\cite{Kadanoff} and \cite{Rammer}, Secs. II A and II B),
\begin{eqnarray}
G^<(\RR,T;\rr,t) &=& +i \langle\langle \psi^+(\rr_2,t_2)\;\psi(\rr_1,t_1)\rangle\rangle,\label{G<}\\
G^>(\RR,T;\rr,t) &=& -i \langle\langle \psi(\rr_1,t_1)  \;\psi^+(\rr_2,t_2)\rangle\rangle.\label{G>}
\end{eqnarray}
$\psi(\rr_1,t_1)$ and $\psi^+(\rr_2,t_2)$
are the electron field operators in the Heisenberg picture.
$\RR = (\rr_1 +\rr_2)/2$ and $T = (t_1 + t_2)/2$
denote a ``center-of-mass" coordinate and a ``mean" time, respectively. In addition relative variables
$\rr=\rr_1-\rr_2$ and $t = t_1 - t_2$ will be needed.
$\langle\langle\cdots\rangle\rangle$  corresponds to the grand-canonical ensemble average (at finite temperatures).

Our starting point is -- as laid out by Sipe and Shkrebtii\cite{Sipe-1} -- an independent particle description with
the Hamiltonian
\begin{equation}
   H(\rr,\pp,t) =  \frac{(\pp - q\AA)^2}{2\;m_0} + V(\rr)+ q\Phi.
   \label{hamiltonian}
\end{equation}
$\pp$ denotes the canonical momentum, $V(\rr)$ is the periodic crystal potential,
and $m_0$ and $q$ are the mass and charge ($q=-e$) of the electrons.
$\AA=\AA(\rr,t)$ and $\Phi=\Phi(\rr,t)$ are the vector and scalar potentials of the
radiation and external (classical) electromagnetic field, $\AA=\AA_{rad}+\AA_{cl}$,  \hbox{div$\AA=0$}.
In the following, we assume that the energies $E_n(\kk)$ and Bloch states $|n,\kk\rangle$ of the electrons
are known from a band structure calculation ($\AA=\Phi=0$).

The photogalvanic effect is independent of photon momentum, see Eq.~(\ref{j}). Therefore, the magnetic field of the
radiation can be neglected; that is, $\AA_{rad}(\rr,t)$ can be approximated by a position-independent
field (equivalent to the electrical dipole approximation), $\AA_{rad}(t)$, $\Phi_{rad}=0$.
Regrouping the remaining terms in Eq.~(\ref{hamiltonian}), we obtain
\begin{eqnarray}
   H(\rr,\pp,t)       &=&  H_{cl}(\rr,\pp,t) + H_{int}(\rr,\pp,t),\label{ham1}\\
   H_{cl}(\rr,\pp,t)  &=&  \frac{(\pp - q\AA_{cl})^2}{2\;m_0} + V(\rr)+ q\Phi_{cl},\label{ham2}\\
   H_{int}(\rr,\pp,t) &=& -\frac{q}{m_0}\big( \pp - q\AA_{cl}(\rr,t)\big) \AA_{rad}(t).\label{ham3}
\end{eqnarray}
Radiation will be treated in terms of a photon propagator; additionally, $\AA_{cl}$ enters as a vertex operator.

In thermal equilibrium ($\AA=\Phi=0$) the Green's functions Eqs.~(\ref{G<},\ref{G>})
can be represented in terms of Bloch functions of Eq.~(\ref{hamiltonian}) ($\AA=\Phi=0$),
\begin{equation}
\mathbf{\hat G}_0 (\RR;\rr,t) =
   \sum_{n,\kk} \langle \RR+\frac{\rr}{2}|n,\kk\rangle \, \mathbf{\hat{g}}_{n,0}(\kk,t) \,\langle n,\kk|\RR-
   \frac{\rr}{2}\rangle,\label{G0}
\end{equation}
where
\begin{equation}
\mathbf{\hat g}_{n,0} =
\left[
	\begin{array}{ll}
	-i\,\theta(t)\; e^{-i\,E_nt} & -i(1-2f_{n,0}(\kk))\; e^{-i\,E_nt} \\
	\,\,\,\,0     &\,\,\,\, i\,\theta(-t)\; e^{-i\,E_nt} \\
	\end{array}
\right].\label{g0}
\end{equation}
$f_{n,0}(\kk)$ denotes the Fermi function.
Here, and in the following, units are used where $\hbar = 1$.

The radiation field will be treated as an external quasiclassical field with no internal dynamics, that is,
there exists only a contribution to the Keldysh component of the photon Green's function $\mathbf{\hat D}$,
%
\begin{equation}
D^K_{\mu\nu}(t) =
 -i\frac{I}{\omega^2\epsilon_0 c\eta} \big( e_{\mu} e^*_{\nu}\; e^{-i\omega t} + cc \big),
 \label{Keldysh-D}
\end{equation}
$cc$ means complex conjugate, for a derivation see Appendix~A.

The equation of motion for $\mathbf{\hat G}$ is identical to the Dyson equation,
\begin{eqnarray}
\mathbf{\hat G}^{-1}_{cl} \otimes\mathbf{\hat G} &=&
        \delta(\rr) \delta(t)\mathbf{\hat 1} + \mathbf{\hat\Sigma}\otimes\mathbf{\hat G},\label{Dyson1}\\
\mathbf{\hat G}^{-1}_{cl} &=& \big( i\,\partial_{t_1} - H_{cl}(\mathbf r_1,\mathbf p_1, t_1)\big)\; \mathbf{\hat 1}\label{Dyson2}.
\end{eqnarray}
$\otimes$ means matrix multiplication, $H_{cl}$ stands for Eq.~(\ref{ham2}),
and $\mathbf{\hat\Sigma}$ denotes  the electron--photon self--energy, which is calculated using $\mathbf{\hat D}$
from Eq.~(\ref{Keldysh-D}), with  $-\frac{q}{m_0} (\pp - q\mathbf A_{cl})$ being the vertex operator (Ref.~\cite{Rammer}, Sec. II C).

\subsection{Kinetic equations}

In order to set up a quasiclassical description the following (standard) approximation for the Green's function with
inclusion of the external electromagnetic field is made, Baym\cite{Baym}~(p. 74)
\begin{widetext}
\begin{equation}
\mathbf{\hat G}(\RR,T;\rr,t) =\sum_{n,n',\kk}\langle \RR+\frac{\rr}{2}|n,\kk\rangle\;\mathbf{\hat g}_{nn'}
(\RR,T;\kk,t)\;\langle n',\kk| \RR-\frac{\rr}{2}\rangle\; e^{iq[\rr\mathbf A_{cl}(\RR,T)- t\Phi_{cl}(\RR,T)]},
\label{G-Approx}
\end{equation}
\end{widetext}
where
\[
\mathbf{\hat g}_{nn'}(\RR,T;\kk,t) =
\left[
	\begin{array}{ll}
	g^R_{nn'} & g^K_{nn'} \\
	\,\,\,\,0     & g^A_{nn'} \\
	\end{array}
\right].
\]
Here,
$I(\RR,T), \mathbf A_{cl}(\RR,T)$ $(\BB=\nabla\times \mathbf A_{cl}(\RR,T))$
and
$\Phi_{cl}(\RR,T)$ $(\EE=-\partial_T\mathbf A_{cl}(\RR,T)-\nabla\Phi_{cl}(\RR,T))$
denote classical macroscopic fields which are assumed to be constant on atomic scales so that Bloch functions are still a
suitable basis and will be noticeable only  in $\mathbf{\hat g}_{nn'}$.
The phase factor $e^{iq\rr\mathbf A_{cl}(\RR,T)}$ takes into account the phase shift induced by a vector potential
$\mathbf A_{cl}$ along the direct path of the particle from $\rr_2$ to $\rr_1$ and reduces the contribution
of the diamagnetic part $q\mathbf A_{cl}$ in the vertex operator $-\frac{q}{m_0} (\pp - q\mathbf A_{cl})$.
Likewise, $e^{-iqt\Phi_{cl}(\RR,T)}$ collects the local shifts of the energy levels due to an electrical potential
$\Phi_{cl}(\RR,T)$.

Observable quantities such as the charge current density $\mathbf j_q$ are calculated with the aid of the Keldysh 
component $ \langle G^K(\RR,t;\rr,t)\rangle$, averaged over the volume of an elementary cell, of Eq.~(\ref{G-Approx}):
\[
\jj_q(\RR,T) = -i\frac{q}{m_0}(\frac{1}{i}\nblr - q\mathbf A_{cl}) \langle G^K(\RR,T;\rr,t)\rangle|_{\rr = 0, t = 0}
\]
where the spin factor of two is already included here.
Using the definition
\hbox{$f_{nn'}(\RR,T;\kk) = \frac{1}{2i}\; g^K_{nn'}(\RR,T;\kk,t=0)$},
the charge current density becomes in terms of $\ff$
\begin{equation}
\jj_q(\RR,T) = \frac{2q}{m_0V}\sum_{n,n',\kk} f_{nn'}(\RR,T,\kk) \langle n',\kk|\pp|n,\kk \rangle\label{ch-cu-den}.
\end{equation}
V is the volume of the crystal.

We are looking for the current contribution which is linear in the intensity (quadratic in the electric field);
therefore, only the ``turtle" photon self--energy diagram is needed.
For the Feynman rules see Ref.\cite{Rammer}~(Eqs.~(2.39-2.43)).
Moreover, only the anti--Hermitian parts of the self-energies $\mathbf{\hat\Sigma}$ (photons and phonons)
will be taken into account because these describe irreversible processes that occur as a consequence of the absorption processes.
Hermitian parts of $\mathbf{\hat\Sigma}$, on the contrary, describe band--renormalization effects which can be
safely neglected\cite{spurious}.

The basic equations for $f_{nn'}$ are obtained from the Dyson equation by subtracting its adjoint, $[\cdots=\cdots]$, and
performing the integral transformation (Ref.\cite{Rammer}, Sec. II E):
\begin{eqnarray*}
&-\frac{1}{2} \int d^3R\int d^3r \,\langle n,\kk |\RR+\frac{\rr}{2}\rangle
  \langle \RR-\frac{\rr}{2}|n',\kk\rangle\\
&\times  e^{-iq[\rr\mathbf A_{cl}(\RR,T)- t\Phi_{cl}(\RR,T)]} \; [\cdots=\cdots].
\end{eqnarray*}
$\RR$-- and $\rr$-- integrations extend over a unit cell and the whole crystal, repectively.
Eventually, the relative time $t$ is set equal to zero. As the result, we obtain:
%
%
\begin{widetext}
\underline{Diagonal elements $f_n = f_{nn}$:}
\begin{equation}
(\partial_T + q\EE\cdot\nblk) f_n(\RR, T;\kk)+ \nblR\cdot\jj_n(\RR, T;\kk) + q\BB\cdot(\nblk\times\jj_n(\RR, T;\kk)) =
G_{n}^{(0)}(\RR, T;\kk) + \delta G_{n}^{(\BB)}(\RR, T;\kk) + I_{n,pn} + I_{n,r}.\label{diagonal}
\end{equation}
This is a modified Boltzmann equation for the distribution function $f_n$ of band n.
The total particle current density $\jj_n(\RR, T;\kk)$ in the drift-  and acceleration terms acts as the driving term,
\begin{equation}
\jj_n(\RR, T;\kk) = \frac{1}{2m_0}\,\sum_{n'}( \langle n,\kk|\pp|n',\kk \rangle\, f_{n'n}(\RR, T;\kk) + Hc) =
 \mathbf v_n(\kk)\, f_n(\RR, T;\kk) + \mathbf j^{ND}_n(\RR, T;\kk),\label{current-density}
\end{equation}
where $Hc$ means Hermitian conjugate. $G_{n}^{(0)}$, $\delta G_{n}^{(\BB)}$, $I_{n,pn}$ and $I_{n,r}$ will be defined below.

In Eq.~(\ref{current-density}), the particle current density is decomposed in terms of a kinetic and a
``nondiagonal" contribution $\jj^{ND}_n$ (see Eq.~(\ref{jnd_n}) below). The latter corresponds to the particle
 shift--current density of the state $\kk$ in the band n and is only different from zero if absorption of  radiation
 causes an interband transition.
\newline
We also obtain:

\underline{Nondiagonal elements $f_{nn'}$ $(n\ne n')$:}
\begin{equation}
i(E_n(\kk) - E_{n'}(\kk))\, f_{nn'}(\RR, T;\kk) =  G_{nn'}^{(0)}(\RR, T;\kk) + \delta G_{nn'}^{(\BB)}(\RR, T;\kk) +
\delta G_{nn'}^{(\EE)}(\RR, T;\kk).\label{nebendiagonal}
\end{equation}
These elements are determined by a comparatively simple equation because there is a dominant term
($i(E_n - E_{n'})\, f_{nn'}$)
on the left side of this equation, in light of which all others
\hbox{($\partial_T f_{nn'}, q\EE\cdot\nblk f_{nn'}$, etc.)}
can safely be neglected.
In order to get a closed set of equations, the particle current density $\jj_n$ and the generation matrix $G_{nn'}$
have still to be specified.
\end{widetext}
The generation matrix  $G_{nn'}(\RR,T;\kk)$ consists of the exclusively intensity dependent part
$G_{nn'}^{(0)}$ with diagonal elements $G_{n}^{(0)} = G_{nn}^{(0)}$
and the parts $\delta G_{nn'}^{(\BB)}$ and $\delta G_{nn'}^{(\EE)}$
which depend linearly on $\BB$ and $\EE$, respectively.
The latter parts stem from the phase factor in Eq.~(\ref{G-Approx}),
and their diagonal elements $\delta G_{n}^{(\BB)}$ and  $\delta G_{n}^{(\EE)}$ are all equal to zero
(dependence on $(\RR, T;\kk)$ is suppressed).
In addition, there is a contribution $\delta G_{n}^{(\BB,dia)}$  from the diamagnetic
part of the vertex operator to Eq.~(\ref{diagonal}) which is exploited in Appendix B.

$I_{n,pn}$ describes the momentum relaxation (e.g., by phonon collisions), and $ I_{n,r}$
describes thermalization and recombination.
As $G_{nn'}$ is a Hermitian matrix, it is conveniently written in the form
\begin{equation}
 G_{nn'}(\kk) = \bar G_{nn'}(\kk) + Hc\label{G-Rate}.
\end{equation}
 There are three contributions to the generation rate $G_{nn'}$:

%
\begin{widetext}
\begin{eqnarray}
&\displaystyle\bar G_{nn'}^{(0)}(\RR,T;\kk) = I(\RR,T)\frac{\pi q^2}{2\,\omega^2\,m_0^2\,\epsilon_0\,c\,\eta}
\sum_{\stackrel{n_1}{\Omega=\pm\omega}}(f_{n_1,0}(\kk) - f_{n',0}(\kk))\;\delta(E_{n_1}(\kk) - E_{n'}(\kk)
 - \Omega) \nonumber\\
&\displaystyle\times\langle n,\kk|p_{\mu}|n_1,\kk \rangle\; \langle n_1,\kk|p_{\nu}|n',\kk \rangle\; e^*_{\mu,\Omega}
\, e_{\nu,\Omega}, \label{Generationsrate}
\end{eqnarray}
%
\begin{eqnarray}
&\displaystyle\delta\bar G_{nn'}^{(\BB)}(\RR,T;\kk) =I(\RR,T)\frac{\pi q^3}{4\,\omega^2\,m_0^2\,\epsilon_0\,c\,\eta}
\sum_{\stackrel{n_1,n_2}{\Omega=\pm\omega}}\Big\{\big[\nabla_{\QQ 1}\times \nabla_{\QQ 2}\big]_\beta \big[(f_{n_2,0}
(\kk+\QQ_2)-f_{n_1,0}(\kk+\QQ_1))\times \big. \Big.\nonumber\\
&\displaystyle \delta(E_{n_2}(\kk+\QQ_2)-E_{n_1}(\kk+\QQ_1)-\Omega)\times\nonumber\\
&\displaystyle\Big. \big.(n,\kk |p_\mu + k_\mu|n_1,\kk +\QQ_1)\; (n_1,\kk +\QQ_1 |p_\nu + k_\nu|n_2,\kk +\QQ_2)  \;
 (n_2,\kk +\QQ_2|n' ,\kk)\;(1-\delta_{n,n'})\big] \Big\}\;  B_\beta \; \frac{1}{i}\; e^*_{\mu,\Omega}\, e_{\nu,\Omega},
 \label{Generationsrate_B}
\end{eqnarray}
%
\begin{eqnarray}
&\displaystyle\delta\bar G_{nn'}^{(\EE)}(\RR,T;\kk) =I(\RR,T)\frac{\pi q^3}{4\,\omega^2\,m_0^2\,\epsilon_0\,c\,\eta}
\sum_{\stackrel{n_1}{\Omega=\pm\omega}}\Big\{\nabla_{\QQ,\alpha} \big[ (f_{n',0}(\kk+\QQ)-f_{n_1,0}(\kk+\QQ))
\times\big.\Big.\nonumber\\
&\displaystyle\partial_{\Omega} \delta(E_{n_1}(\kk+\QQ)-E_{n'}(\kk+\QQ)+\Omega)\times\nonumber\\
&\displaystyle\Big. \big.(n,\kk|p_\mu + k_\mu|n_1,\kk +\QQ)\; (n_1,\kk +\QQ|p_\nu + k_\nu|n',\kk )\big]\Big\} \;  E_\alpha \;
\frac{1}{i}\; e^*_{\mu,\Omega}\, e_{\nu,\Omega}. \label{Generationsrate_E}
\end{eqnarray}
After differentiation, the vectors $\QQ$, $\QQ_1$, and $\QQ_2$ have to be set to zero. The expressions
$(n_1,\kk_1| . . . |n_2,\kk_2)$ are matrix elements, which are calculated with respect to the lattice--periodic parts of
the Bloch functions, and $e_{\mu,\omega} = e_{\mu}$ and $e_{\mu,-\omega} = e^*_{\mu}$ are the  components of
 the complex--valued polarization vector.

$\mathbf j^{ND}_n$ is obtained from Eq.~(\ref{current-density}) with $\bar G_{nn'}^{(0)}$ from
 Eq.~(\ref{Generationsrate}):
%
%
\begin{eqnarray}
&\displaystyle j_{n,\alpha}^{ND}(\RR, T;\kk) = \frac{1}{m_0}\;\sum_{m\ne n}\Im\left (\frac{
\langle n,\kk|p_{\alpha}|m,\kk \rangle\; \bar G_{mn}^{(0)}(\RR,T;\kk) +  \bar G_{nm}^{(0)*}(\RR,T;\kk)\;
\langle m,\kk|p_{\alpha}|n,\kk\rangle^*}{E_m-E_n}\right )\label{jnd_n}\\
&\displaystyle=I(\RR,T)\frac{\pi e^2}{2\,\omega^2\,m_0^3\,\epsilon_0\,c\,\eta}\;\sum_{\stackrel{m\ne n,n_1}
{\Omega =\pm\omega}}\big[(f_{n_1,0}-f_{n,0})\;\delta(E_{n_1}-E_n-\Omega)+(f_{n_1,0}-f_{m,0})\;
\delta(E_{n_1}-E_m-\Omega)\big]\label{jnd}\\
&\displaystyle\times\left [\Im\left(\frac{\langle n,\kk|p_{\alpha}|m,\kk\rangle\langle m,\kk|p_{\mu}|n_1,\kk\rangle
\langle n_1,\kk|p_{\nu}|n,\kk\rangle}{E_m-E_n}\right)\; \Re (e^*_{\mu,\Omega}\, e_{\nu,\Omega})\right.+
 \label{Teil1}\\
&\displaystyle \quad\left. \Re\left(\frac{\langle n,\kk|p_{\alpha}|m,\kk\rangle\langle m,\kk|p_{\mu}|n_1,\kk\rangle
\langle n_1,\kk|p_{\nu}|n,\kk\rangle}{E_m-E_n}\right)\; \Im (e^*_{\mu,\Omega}\, e_{\nu,\Omega})\right ].
\label{Teil2}
\end{eqnarray}
The term~(\ref{Teil1}) is an even function of $\kk$ that contributes to $\PP^S$, Eq.~(\ref{Pxyz}),
whereas the odd term~(\ref{Teil2}) does not.
\end{widetext}
%
\section{Derivation of the PG tensors}

As an application of the kinetic theory we verify the result Eq.~(\ref{shiftcur}) for $\PP^S$ and give the representations
of the other PG tensors $\PP^A$, $\RR^S$, and $\RR^A$ as defined by Eq.~(\ref{j}).
The following assumptions are made:
(i) there is no external electrical field,
(ii) there is  an external magnetic field $\BB$, and the (monochromatic) radiation intensity $I$ is constant in space and time
so that  $f_{nn'}$ does not depend on $(\RR,T)$.
Under these assumptions the kinetic equations~(\ref{diagonal}-\ref{nebendiagonal}) become
\begin{widetext}
\begin{eqnarray}
n=n':\quad       && q\;\BB\cdot(\nabla_k\times\jj_n(\kk)) =
                    G_{n}^{(0)}(\kk) - \frac{f_n(\kk)-\langle f_n(\kk)\rangle_E }{\tau_n} + I_{n,r},\label{diag}\\
n\not= n':\quad  && i\;(E_n(\kk) - E_{n'}(\kk)) \;  f_{nn'}(\kk)
                    = G_{nn'}^{(0)}(\kk) + \delta G_{nn'}^{(\BB)}(\kk).\label{nebendiag}
\end{eqnarray}
In addition, Eqs.~(\ref{current-density}) and (\ref{jnd_n}--\ref{Teil2}) will be needed.
\end{widetext}

To simplify matters, the collision operator $I_{n,pn}$ was replaced within a relaxation time approximation.
$\langle f_n(\kk)\rangle_E$ denotes the average of the distribution function over a surface of constant energy,
and $\tau_n$ is the relaxation time for each band $n$; numerical values are taken from experiment.
The operator $I_{n,r}$ which ensures thermalization and recombination is assumed to be only  energy dependent.
Obviously, the PG current solely stems from $f_{nn'}(\kk)$ terms, which are asymmetric  with respect to $\kk$,
\hbox{$\delta f_{nn'}(\kk) = -\delta f_{nn'}^*(-\kk)$},
which in turn originate from generation terms with
\hbox{$\delta G_{n}(\kk) =  -\delta G_{n}(-\kk)$}
and
\hbox{$\delta G_{nn'}(\kk) = \delta G_{nn'}^*(-\kk)$}.
Therefore, only such terms will be considered when deriving representations for the tensors.

\subsection{Tensor $\PP^S$}
Linearly polarized light and $\BB=0$ are implied in Eqs.~(\ref{diag},\ref{nebendiag}).
The relevant contributions of the state function are:
\begin{eqnarray*}
n=n':\quad && \delta f_n = 0,\\
n\not= n':\quad &&\delta f_{nn'} = \frac{G_{nn'}^{(0)}(\kk)}{i\; (E_n(\kk) - E_{n'}(\kk))}.
\end{eqnarray*}
%

The corresponding PG current density $\mathbf j_{\mbox{\tiny PG}}$ is determined from Eq.~(\ref{jnd_n}) by
summation over all states (including the spin factor of two)
\begin{eqnarray*}
\mathbf j_{\mbox{\tiny PG}} &=& \frac{2q}{V} \sum_{n, \kk} \jj_{n}^{ND}(\kk),
\end{eqnarray*}
which is performed along the route described in Refs.\cite{RvB-3,Kristoffel-1}.
As a result, we obtain\cite{twobands}:
\begin{widetext}
\begin{eqnarray}
&& P^S_{\alpha\mu\nu} =
   \frac{e^3}{4\pi^2\,\omega^2\,m_0^2\,\epsilon_0\,c\,\eta} \int\limits_{1.BZ} d^3k\,(f_{v,0}-f_{c,0})\;
     \delta ( E_c(\kk)-E_v(\kk)-\omega ) \times \nonumber \\
&& \Big \{ \frac{1}{2}\;\Im\big [ ( \nabla_{\kk,\alpha} \langle c,\kk \left| p_{\nu} \right| v,\kk \rangle )\;
\langle v,\kk \left| p_{\mu} \right| c,\kk \rangle -   \langle c,\kk \left| p_{\nu} \right| v,\kk \rangle
 ( \nabla_{\kk,\alpha}\langle v,\kk \left| p_{\mu} \right| c,\kk \rangle )\big ]\nonumber\\
&&  \,\phantom{\{}+ \,\Re\big [ \langle c,\kk \left| p_{\nu} \right| v,\kk \rangle\;
   \langle v,\kk \left| p_{\mu} \right| c,\kk \rangle ] \; [ X_{vv,\alpha} - X_{cc,\alpha}\big ] \Big\}.
   \label{Pxyz}
\end{eqnarray}
Equation~(\ref{Pxyz}) is identical to Eq.~(\ref{shiftcur}), as can be checked by decomposing $\ee\cdot\pp$
 into components.
%
\subsection{Tensor $\PP^A$}
Circularly polarized light and $\BB=0$ are implied in Eqs.~(\ref{diag},\ref{nebendiag}).
The relevant contribution of the state function to determine  $\PP^A$  is:
\begin{eqnarray}
n=n':\quad && \delta f_n = \tau_n \;\delta G_{n}^{(0)}(\kk),\label{df-circular}\\
n\not= n':\quad &&\delta f_{nn'} = 0.
\end{eqnarray}
$\delta G_{n}^{(0)}(\kk)$ is the part of the generation rate $G_{n}^{(0)}(\kk)$ as given by
Eqs.~(\ref{G-Rate},\ref{Generationsrate}),
%
\begin{eqnarray}
&\displaystyle\delta G_{n}^{(0)}(\kk) = I \frac{\pi q^2}{\omega^2\,m_0^2\,\epsilon_0\,c\,\eta}
\sum_{\stackrel{n'}{\Omega=\pm\omega}}(f_{n,0}(\kk) - f_{n',0}(\kk))\;\delta(E_{n'}(\kk) - E_{n}(\kk) - \Omega)\nonumber\\
&\displaystyle\times\Im\big (\langle n,\kk|p_{\mu}|n',\kk \rangle\; \langle n',\kk|p_{\nu}|n,\kk \rangle\big )\;
\Im(e^*_{\mu,\Omega}\, e_{\nu,\Omega}). \label{G-Rate-circular}
\end{eqnarray}
Insertion of Eq.~(\ref{df-circular}) and Eq.~(\ref{G-Rate-circular}) into Eq.~ (\ref{ch-cu-den}) leads to
\begin{equation}
\displaystyle j_{\alpha}^{circ} = \frac{2q}{V} \sum_{\stackrel{n=v,c}{\kk}} v_{n,\alpha}(\kk)\;\delta f_n(\kk),
\label{j_circ}
\end{equation}
and the tensor element $P_{\alpha\mu\nu}^A$ reads:
\begin{eqnarray}
&\displaystyle P_{\alpha\mu\nu}^A = \frac{e^3}{4\pi^2\,\omega^2\,m_0^2\,\epsilon_0\,c\,\eta}
\sum_{\stackrel{n',n}{\Omega=\pm\omega}}\int\limits_{\,1.BZ}\!d^3k\,\,(f_{n',0}(\kk) - f_{n,0}(\kk))\;
\delta(E_{n'}(\kk) - E_{n}(\kk) - \Omega) \nonumber\\
&\displaystyle\times\tau_n\; v_{n,\alpha}(\kk)\;(\delta_{v,n}+\delta_{c,n})\;\Im(\langle n,\kk|p_{\mu}|n',\kk \rangle
\; \langle n',\kk|p_{\nu}|n,\kk \rangle)\;\sign(\Omega).\label{PA}
\end{eqnarray}
Performing the sums over $\Omega$, n and n', we obtain\cite{twobands}:
\begin{eqnarray}
&\displaystyle P_{\alpha\mu\nu}^A = \frac{e^3}{4\pi^2\,\omega^2\,m_0^2\,\epsilon_0\,c\,\eta}\int\limits_{\,1.BZ}
\!d^3k\,\,(f_{v,0}(\kk) - f_{c,0}(\kk))\;\delta(E_{c}(\kk) - E_{v}(\kk) - \omega) \nonumber\\
&\displaystyle\times\Big (\tau_c\; v_{c,\alpha}(\kk) - \tau_v\; v_{v,\alpha}(\kk)\Big )\;\Im\Big (
\langle v,\kk|p_{\mu}|c,\kk \rangle\; \langle c,\kk|p_{\nu}|v,\kk \rangle \Big).\label{PA2}
\end{eqnarray}
In contrast to the linear PGE ($\PP^S$ term) the circular PGE is ballistic as only diagonal elements of the state function
contribute and it depends on the scattering times of the (hot) photo--generated carriers.
\end{widetext}
\subsection{Tensor $\RR^S$ }
Linearly polarized light and $\BB\ne 0$ are implied in Eqs.~(\ref{diag},\ref{nebendiag}). The relevant contributions of
the state function are:
\begin{eqnarray}
n=n':\quad && \delta f_n = -q\,\tau_n\;\BB\cdot\big[\nblk\times\jj_n(\kk)\big],\label{dfB}\\
n\not= n':\quad &&\delta f_{nn'} = \frac{G_{nn'}^{(0)}(\kk)}{i\; (E_n(\kk) - E_{n'}(\kk))}.\label{dfnB}
\end{eqnarray}
The first equation describes the portion of the charge current density which is deflected by the magnetic field,
analogous to the Hall effect. The current density $\jj_n(\kk)$ is inserted from Eq.~(\ref{current-density}) and the
contribution of Eq.~(\ref{dfnB}) is used therein as the driving term.
The resulting charge current density $\jj^{\mbox{\tiny Hall}}$ reads:
\begin{widetext}
\begin{eqnarray}
\displaystyle j_{\alpha}^{Hall} &=& \frac{2q}{V} \sum_{\stackrel{n=v,c}{\kk}} v_{n,\alpha}(\kk)\;\delta f_n(\kk)
 =\frac{-2q^2}{V}\,\, B_{\beta}\;\epsilon_{\beta \gamma \delta} \sum_{\stackrel{n=v,c}{\kk}} v_{n,\alpha}(\kk)
 \;\tau_n\; \nabla_{\kk,\gamma}\, j^{ND}_{n,\delta}(\kk),\nonumber\\
 \displaystyle j_{\alpha}^{Hall}&=& \frac{2q^2}{V}\,\, B_{\beta}\; \epsilon_{\beta \gamma \delta}
 \sum_{\stackrel{{n=v,c}}{\kk}} \nabla_{\kk,\gamma}\big (v_{n,\alpha}(\kk)\; \tau_n\big )\; j^{ND}_{n,\delta}(\kk).
 \label{j-hall}
\end{eqnarray}
Inserting Eqs.~(\ref{jnd},\ref{Teil1}) into Eq.~(\ref{j-hall}), we get for the tensor element $R_{\alpha\beta\mu\nu}^S$
the expression:
\begin{eqnarray}
&\displaystyle R^S_{\alpha\beta\mu\nu}=\frac{e^4}{16\pi^2\,\omega^2\,m_0^3\,\epsilon_0\,c\,\eta}\;
\epsilon_{\beta\gamma\delta}\sum_{\stackrel{m\ne n,n_1
}{n,\Omega=\pm\omega}}\int\limits_{\,1.BZ}\!d^3k\,\,\big (f_{n_1,0}(\kk) - f_{n,0}(\kk)\big )\;\delta(E_{n_1}(\kk)
 - E_{n}(\kk) - \Omega)\times\nonumber\\
&\displaystyle\big(\nabla_{\kk,\gamma}\big[\tau_n v_{n,\alpha}(\kk)(\delta_{v,n}+\delta_{c,n}) + \tau_{m}v_{m,\alpha}
(\kk)(\delta_{v,m}+\delta_{c,m})\big]\big)\times\nonumber\\
&\displaystyle\Im\left(\frac{\langle n,\kk|p_{\delta}|m,\kk\rangle\langle m,\kk|p_{\mu}|n_1,\kk\rangle
\langle n_1,\kk|p_{\nu}|n,\kk\rangle}{E_m-E_n}
+ \textnormal{terms with }\mu\textnormal{ and }\nu\textnormal{ interchanged}\right).\label{Rwxyz}
\end{eqnarray}
Performing all sums\cite{twobands} leads to $\RR^S$:
\begin{eqnarray}
&\displaystyle R^S_{\alpha\beta\mu\nu}=\frac{e^4}{16\pi^2\,\omega^2\,m_0^2\,\epsilon_0\,c\,\eta}\;
\epsilon_{\beta\gamma\delta}\int\limits_{1.BZ}\!d^3k\,\,(f_{v,0}(\kk) - f_{c,0}(\kk))\;
\delta(E_{c}(\kk) - E_{v}(\kk) - \omega)\times\nonumber\\
&\displaystyle\bigg \{\big (\nabla_{\kk,\gamma}\; \tau_c v_{c,\alpha}(\kk)\big )\;
\left[- \Im (\langle v,\kk \left|p_{\nu}\right|c,\kk\rangle\;
             \langle c,\kk\left |{\mathcal{R}}^{\dagger}_{\delta}\, p_{\mu}\right|v,\kk\rangle )
  + \frac{v_{c,\mu}}{\omega}\;
 \Im ( \langle v,\kk|p_{\nu}|c,\kk\rangle\;\langle c,\kk|p_{\delta}|v,\kk\rangle ) \right ]  \bigg.\nonumber \\
&\displaystyle\bigg. + \big (\nabla_{\kk,\gamma}\; \tau_v v_{v,\alpha}(\kk)\big )\; \left [-\Im
 (\langle v,\kk \left|p_{\nu}\right|c,\kk\rangle\;\langle c,\kk\left |p_{\mu}{\mathcal{R}}_{\delta}\right|v,\kk\rangle)
  - \frac{v_{v,\mu}}{\omega}\;
 \Im ( \langle v,\kk|p_{\nu}|c,\kk\rangle\;\langle c,\kk|p_{\delta}|v,\kk\rangle )\right ]  \bigg. \nonumber\\
&\big. \textnormal{plus all terms with } \mu \textnormal{ and } \nu \textnormal { interchanged} \bigg \}.
\label{RSwxyz}
\end{eqnarray}
\end{widetext}
${\RRR}$ is the shift operator\cite{RvB-3,Kristoffel-1}, in position representation
\[
{\RRR}_{n,\kk}(\rr) = \langle\rr\left| {\RRR} \right| n,\kk \rangle = e^{i\kk\rr}
\left\{ \nblk + i \XX_{nn}(\kk) \right\} u_{n\kk}(\rr).
\]
The shift operator ${\RRR}$ is of importance when photogalvanic current densities are described by the nondiagonal
elements of the state function $\ff$. In particular, the shift vector Eq.~(\ref{shiftvec}) can be expressed as
\begin{equation}
\mathbf{s}_{cv}(\ee,\kk) = \frac{\Im\big(\langle c,\kk\left |{\RRR}^{\dagger}\,\ee\pp+\ee\pp \, {\RRR}\right|v,\kk\rangle\;
\langle v,\kk \left|\ee\pp\right|c,\kk\rangle\big)}{\langle v,\kk \left|\ee\pp\right|c,\kk\rangle\;
\langle c,\kk \left|\ee\pp\right|v,\kk\rangle}.
\end{equation}
The elements of $\RR^S$ show almost the same \hbox{$\omega$}  dependence as those of $\PP^S$ and as a rule of thumb,
\hbox{$|R^S| \approx |P^S|\cdot \mu$} may be expected, where $\mu$  is the mobility of the (hot) photocarriers.

%
%
Result (\ref{RSwxyz}) is completed by the diamagnetic contribution Eq.~(\ref{dia-rs})
\[
\displaystyle R^{S,dia}_{\alpha\beta\mu\nu} = \frac{e}{\omega\, m_0}\,
\epsilon_{\beta\nu\gamma}\, P_{\alpha\mu\gamma}^A.
\]

\subsection{Tensor $\RR^A$ }
Circularly polarized light and $\BB\not=0$ are implied in Eqs.~(\ref{diag},\ref{nebendiag}).
The relevant contributions are
%
\begin{eqnarray}
n=n':\,\, && \delta f_n = -q\tau_n\BB\cdot\big[\nabla_k\times (\tau_n\vv_n\delta G_{n}^{(0)})\big],\label{ballistic}\\
n\not= n':\,\, &&\delta f_{nn'} = \frac{\delta G_{nn'}^{(\BB)}(\kk)}{i\; (E_n(\kk) - E_{n'}(\kk))}.\label{shift}
\end{eqnarray}

Equation~(\ref{ballistic}) describes the deflection of the ballistic charge current density Eq.~(\ref{j_circ}) by the
magnetic field and is present only -- like $\PP^A$ -- in gyrotropic media, whereas  the contribution Eq.~(\ref{shift})
is directly related to the the change oin the generation matrix by the external magnetic field $\BB$.
Therefore, $\RR^A$ consists of two contributions,
\begin{equation}
\RR^A = \RR^{A,\textnormal{bal}} + \RR^{A,\textnormal{shift}}.
\end{equation}

\begin{widetext}
\subsubsection{\textnormal{\textbf{Tensor}} $\RR^{A,\textnormal{bal}}$}
Equation~(\ref{ballistic}) is equivalent to Eq.~(\ref{dfB}).
Following the same route as taken by Eqs.~(\ref{j-hall},\ref{RSwxyz}) and using Eq.~(\ref{G-Rate-circular}),
we arrive at\cite{twobands}
\begin{eqnarray}
&\displaystyle R_{\alpha\beta\mu\nu}^{A,bal} = \frac{e^4}{4\pi^2\,\omega^2\,m_0^2\,\epsilon_0\,c\,\eta}
\;\epsilon_{\beta\gamma\delta}\int\limits_{\,1.BZ}\!d^3k\,\,\big(f_{v,0}(\kk) - f_{c,0}(\kk)\big)\;
\delta(E_{c}(\kk) - E_{v}(\kk) - \omega)\times \nonumber\\
&\displaystyle\Big (\tau_v\; v_{v,\delta}(\kk)\;\big(\nblk_{,\gamma}\tau_v  v_{v,\alpha}(\kk)\big)-\tau_c
\; v_{c,\delta}(\kk)\; \big(\nblk_{,\gamma}\tau_c v_{c,\alpha}(\kk)\big)\Big )\;\Im\Big (
\langle v,\kk|p_{\mu}|c,\kk \rangle\; \langle c,\kk|p_{\nu}|v,\kk \rangle\Big).\label{RAbal}
\end{eqnarray}
\subsubsection{\textnormal{\textbf{Tensor}} $\RR^{A,\textnormal{shift}}$}
 The corresponding current density is
\begin{eqnarray}
j_{\alpha}^{ND,\BB} = \frac{4q}{m_0} \sum_{\stackrel{n,n'}{n\not=n'}}\frac{1}{(2\pi)^3}\int\limits_{1.BZ}\!
\Im\left(\frac{\langle n,\kk|p_{\alpha}|n',\kk \rangle\;\delta\bar  G_{n'n}^{(\BB)}(\kk)}{E_{n'}-E_n}\right)
\,d^3k. \label{Stromdichte_B_shift}
\end{eqnarray}
Inserting $\delta\bar G_{n'n}^{(\BB)}(\kk)$ from Eq.~(\ref{Generationsrate_B}) and regrouping terms we get
\begin{eqnarray}
&\displaystyle j_{\alpha}^{ND,\BB} = I\; \frac{e^4}{16\pi^2\,\omega^2\,m_0^3\,\epsilon_0\,c\,\eta}
\sum_{\stackrel{n_1,n_2}{\Omega=\pm\omega}}\int\limits_{1.BZ}\!d^3k \big\{ \big[\nabla_{\QQ 1}\times
 \nabla_{\QQ2}\big]_{\beta}\; \big[ \big (f_{n_2,0}(\kk + \QQ_2) - f_{n_1,0}(\kk + \QQ_1)\big )\big.\big.\nonumber\\
&\big. \big.\times\delta(E_{n_2}(\kk+\QQ_2) - E_{n_1}(\kk+ \QQ_1) - \Omega) \;\sign(\Omega ) \;
 M^{n_1 n_2}_{\alpha\mu\nu}(\kk,\QQ_1,\QQ_2)\big] \big\} \; B_{\beta}\; \Im(e^*_\mu e_{\nu}),\label{Stromdichte_B}
\end{eqnarray}
with
\begin{eqnarray}
 \displaystyle\lefteqn{ M^{n_1 n_2}_{\alpha\mu\nu}(\kk,\QQ_1,\QQ_2) =}\nonumber\\
& &\displaystyle\sum_{\stackrel{n,n'}{n\not =n'}} \Im \Big \{ (n,\kk|n_2 ,\kk+\QQ_2)\;
(n_2,\kk+\QQ_2 |p_\nu + k_\nu|n_1,\kk +\QQ_1)\; \frac{ (n_1,\kk +\QQ_1 |p_\mu + k_\mu|n',\kk)\;
 (n',\kk|p_{\alpha} + k_{\alpha} |n,\kk)}{E_{n} - E_{n'}}\Big.\nonumber\\
& &\Big. \textnormal{minus all terms with } \mu \textnormal{ and } \nu \textnormal { interchanged} \Big \},\nonumber\\
& &\displaystyle= m_0\;\sum_n \Im \Big \{ (n,\kk|n_2 ,\kk+\QQ_2))\;
(n_2,\kk+\QQ_2 |p_\nu + k_\nu|n_1,\kk +\QQ_1)\; (n_1,\kk +\QQ_1 |(p_\mu + k_\mu)\; {\mathcal{R}}_\alpha |n,\kk)\Big.
\label{Im_R}\\
& & \big. \textnormal{minus all terms with } \mu \textnormal{ and } \nu \textnormal{ interchanged} \Big \}.\nonumber
\end{eqnarray}
In expression~(\ref{Im_R}) we have used the representation of the shift operator ${\RRR}$ with respect to the
lattice--periodic part of the Bloch functions,
$(\rr|{\RRR}|n,\kk) = (\nblk + i \XX_{nn}(\kk)) u_{n\kk}(\rr)$.

As a result, we obtain\cite{twobands}
\begin{eqnarray}
&\displaystyle R^{A,shift}_{\alpha\beta\mu\nu}=\frac{e^4}{16\pi^2\,\omega^2\,m_0^3\,\epsilon_0\,c\,\eta}
\int\limits_{1.BZ} \!d^3k \big\{  \big[\nabla_{\QQ 1}\times \nabla_{\QQ2}\big]_{\beta}\;
\big[\big (f_{v,0}(\kk + \QQ_1) - f_{c,0}(\kk + \QQ_2)\big )\big. \big.\nonumber\\
&\big.\big.\times\delta(E_{c}(\kk +  \QQ_2) - E_{v}(\kk + \QQ_1) - \omega) \;
 \big (M^{cv}_{\alpha\mu\nu}(\kk,\QQ_2,\QQ_1) -
 M^{vc}_{\alpha\mu\nu}(\kk,\QQ_1,\QQ_2)\big)\big]\big\}.\label{RAshift}
\end{eqnarray}
After differentiation, the vectors $\QQ_1$ and $\QQ_2$ have to be set to zero.
\end{widetext}
Due to the differentiations with respect to $\QQ_1$  and $\QQ_2$, even an approximate evaluation of  the tensor elements
of $\RR^A$ requires details of the band structure $E_n(\kk)$ and momentum matrix elements, at least at
a symmetry point $\kk_0$ where the optical transition occurs. If the bands are isotropic near $\kk_0$, the cross--product
operation
$\big (\nabla_{\QQ 1}\times \nabla_{\QQ2}\big)$
whose terms are exclusively dependent on $\QQ_i$ via the energy $E(\kk+\QQ_i)$, does not contribute.
We therefore expect warped energy bands as a favorite ingredient for the circular shift magneto--PGE.

\section{Summary and Discussion}
We have developed  a systematic semiclassical description of the PGE within the Kadanoff--Baym--Keldysh technique
which ensures gauge invariance as well as particle conservation from the beginning.
In addition, band--renormalization terms (Hermitian parts of the self-energies $\mathbf{\hat\Sigma}$) are identified,
and external (slowly varying) electric and magnetic fields are included.
This approach is based on a Boltzman--type equation for the diagonal elements of the state operator and captures nondiagonal
contributions by simple algebraic equations, similar to the well--known
semiconductor--Bloch--equations\cite{HaugKoch} (but without Coulomb interaction).

In our approach, the PGE is a band structure property of the noncentrosymmetric crystal, and the photogalvanic current
is caused by the absorption of light in combination with (symmetric) scattering by phonons and impurities.
Sections~III A--III D gave explicit results for the tensors $\PP^S,\PP^A,\RR^S$ and $\RR^A$.
Here, only the case of an external magnetic field was considered because the influence of an electrical field on the PGE
was studied recently in detail by Fregoso\cite{Fregoso}.
Not included are (i) asymmetric scattering terms, (ii) the magnetic field dependence of scattering,
and (iii) transitions from bound impurity states.
Result (\ref{Pxyz}) for $\PP^S$ is identical to the known result of Eq.(\ref{shiftcur}) and serves as a check,
 whereas results for $\PP^A$, $\RR^S$, and $\RR^A$ are new.
Here, $\PP^A$, Eq.(\ref{PA2}), is equivalent to Eq.(29) of Ref.\cite{Yan,Yan-private}.
Implementation of the p--matrix elements within density functional theory (DFT) calculations is described in Ref.\cite{abinitio}.
Appendix C provides a numerical application to GaP.

For linear polarization there are several examples which clearly demonstrate that the magnitude and spectral structure
are dominated by the shift mechanism:
(i) n--GaP\cite{Hornung-1} (pseudopotential theory) and
(ii) BaTiO$_3$\cite{Rappe-1,Rappe-3} (DFT includes the calculated phonon spectrum and electron--phonon couplings).
In both cases, there is almost perfect agreement with experiment\cite{Gibson,Wuerfel};
nevertheless, asymmetric phonon contributions cannot be excluded in general.
For GaAs a purely ballistic theory gave a good overall description, but the predicted spectrum differed from that
observed\cite{Alperovich}.
For a critique of the shift mechanism as a main source of the PGE see Sturman\cite{Sturman-2}.

Nonlocal aspects of the PGE are usually neglected but have shown up in connection with the analysis of
volume--phase holograms in ferroelectrics\cite{hologram}.
Such phenomena are captured by the semiclassical description, Eqs.~(\ref{diagonal}--\ref{nebendiagonal}),
and may become relevant for optical nano--devices, as recently studied by local photoexcitation\cite{Nakamura},
and are under discussion in connection with spatiotemporal quantum pumping by femtosecond light pulses\cite{Bajpai}.

Quantum kinetic descriptions for the PGE were implicitly used in several previous publications,
e.g. Belinicher et al.\cite{Belinicher-2},
Deyo et al.\cite{Deyo} worked out a semiclassical theory of nonlinear transport and the PGE
but only the influence of electric and magnetic fields on the scattering probabilities were considered, and recently,
Kral\cite{Kral-2} presented a quasiclassical description of the PGE for the problem of electron pumping in semiconductors.
Barik and Sau\cite{Barik} showed that the PGE/BPVE can be attributed to the dipole moment
of the photogenerated excitons, which resembles the difference $[ X_{vv,\alpha} - X_{cc,\alpha} ]$ in Eq.~(\ref{Pxyz}).
The first attempt, probably, for a systematic theory in terms of the Kadanoff--Baym--Keldysh technique
was undertaken by one of the present authors (D.H.) in Ref.\cite{Hornung-Diss}.

There are several numerical studies of the shift vector ${\mathbf s}_{cv}(\ee,\kk)$ as well as an analytic estimate
to find optimal parameters (concerning  band structure and polarization directions) for the PG response\cite{Rappe-1,Rappe-2}.
These investigations, however, are based on a simplified version of the shift vector Eq.~(\ref{shiftvec}) with restricted
combinations of the current and light--polarization components (see discussion around Eq.~(58) in Ref.\cite{Sipe-1}).
To overcome such restrictions, we have worked out the general coordinate--free form of  the shift vector given by
Eqs.~(\ref{shiftcur}--\ref{Berry-X}).

In an external magnetic field $\BB$, the currents described by $\PP^S$ and $\PP^A$ are deflected like Hall currents,
which result in ballistic contributions described by $\RR^S$ (proportional to the mobility) and 
$\RR^{A,\textnormal{bal}}$ [proportional to the square of the mobility; see Eqs.~(\ref{dfnB}) and (\ref{ballistic})].
In addition, $\RR^A$ includes a shift contribution $\RR^{A,\textnormal{shift}}$, which is related to the influence of
magnetic field $\BB$ on the generation matrix $G_{nn'}(\kk)$.
Concerning the experimental situation, we refer to the work of Fridkin and his group, see Refs.~\cite{Fridkin,Sturman-1}.
For tellurium theoretical and experimental studies are due to Ivchenko et al.\cite{Ivchenko-2,Ivchenko-4}.
However, application of their theoretical results in first--principles calculations does not seem to be straightforward.

The Hall property of the linear PGE in a magnetic field (described by $\RR^S$) has been used to determine the mobility
of photogenerated charge carriers\cite{Alperovich,Fridkin,Zong}.
Very large mobilities have been reported: $0.5\times 10^6$cm$^2$/Vs (4.2K) for GaAs,
 approximately $6000$cm$^2$/Vs for piezoelectric Bi$_{12}$GeO$_{20}$ (point group 23), 
 and up to $1900$cm$^2$/Vs (room temperature) for ferroelectric BaTiO$_3$ (point group 4mm).
The analysis of the measurements is based on the standard Hall formula,
\begin{equation}
 \jj^{Hall}=\mu\;\jj^{(0)}\times\BB,
 \label{Hallformula}
\end{equation}
which stems from a Drude--type description and holds under isotropic conditions.
For Bi$_{12}$GeO$_{20}$ the PG current without magnetic field $\jj^{(0)}$ is strongest just below the gap ($3.2$eV)
and is believed to originate from impurity transitions into the conduction band;  that is, it is of ballistic type.
Hence, Eq.~(\ref{Hallformula}) is a suitable basis for the experimental analysis.
For BaTiO$_3$, however, the PGE is mainly due to interband transitions\cite{Rappe-1,Rappe-3,Wuerfel},
 so that Eq.~(\ref{Hallformula}) is not appropriate, even if $\mu_c\gg\mu_v$, compare Eq.~(\ref{Pxyz}) with (\ref{RSwxyz}).

The idea to separate shift and ballistic contributions of the PG current by using a magnetic field in combination
with linearly and circularly polarized light has been pursued by Fridkin and collaborators, see e.g. Ref.\cite{Fridkin}
and, more recently, by Burger et al.\cite{Burger-1,Burger-2} for Bi$_{12}$GeO$_{20}$ and Bi$_{12}$SiO$_{20}$.
Their analysis, however, is based on the assumption that the shift mechanism does not contribute to the photo Hall current
(``$j_{sh}$ describes coherence between wave packets rather than a transport process", see above Eq.~(1) of Ref.~\cite{Burger-2}),
which is at odds with our results as given by Eqs.~(\ref{j-hall}) and (\ref{Stromdichte_B_shift}).
It  also contradicts a previous result of Ref.\cite{Ivchenko-4} (their formula (13)).
Moreover, in these studies the PG current is due to ("ballistic") impurity transitions and does not originate from
interband transitions, which are the origin of the shift mechanism\cite{comment-2}.

\begin{acknowledgments}
  We thank Peter W\"olfle for his advice and support with the preparation of the manuscript and
  Binghai Yan and Zhenbang Dai for discussions.
\end{acknowledgments}

\appendix
\section{Photon Green's function}
The Keldysh Green's function $\mathbf{D}_{\mu\nu}$ for photons has the usual Jordan normal form, and each
matrix element is a polar tensor of rank two. We start from (Ref.\cite{Rammer},~Sec.~IIA)
\begin{equation}
D_{\mu\nu}^< (\rr_1,t_1;\rr_2,t_2)= -i \; \langle\langle A_\nu(\rr_2,t_2)\;A_\mu(\rr_1,t_1)\rangle\rangle.
\label{A1}
\end{equation}
$A_\mu(\rr_j,t_j)$ ($j=1,2$) denotes the (Hermitian) vector potential (field operator) of the radiation and $\mu$ and $\nu$
refer to the polarization of the photons.  $D_{\mu\nu}^{>}(\rr_1,t_1;\rr_2,t_2)=D_{\nu\mu}^{<}(\rr_2,t_2;\rr_1,t_1)$;
the other photon--correlation functions are defined in the same way as for the electrons.

As thermal radiation at ambient temperature plays no role, radiation is described as a classical external field
of a single mode. Its quantum analog is a coherent state $|\alpha\rangle$, $a|\alpha\rangle=\alpha|\alpha\rangle$,
$\langle\langle\dots\rangle\rangle \to \langle\alpha| \dots |\alpha\rangle$.
$a$ and $a^\dagger$ denote destruction and creation operators of the mode,  $a a^{\dagger} - a^{\dagger} a = 1$.
$\alpha=|\alpha| \exp(i\phi)$ is a complex number, where $|\alpha|^2$ is the mean photon number of the mode which is
proportional to the light intensity.

The vector potential operator reads
\[ A_\mu (\rr_j,t_j) = \;\sqrt{\frac{1}{2\epsilon\epsilon_0 V}} \frac{1}{\omega}
 \big(\;e_\mu \; a \; e^{{i (\qq\rr_j - \omega t_j)}} + Hc \; \big), \]
where $\qq$, $\omega=\omega(\qq)$, and $\ee$ denote the wave vector, frequency, and polarization vector of the mode.
$\epsilon = \eta^2$ is the dielectric constant of the  medium, and $V$ is the volume of the
cavity (periodic boundary conditions are implied), see e.g. (Louisell\cite{Louisell},~Sec.~4.3).
To simplify notation mode indices have been suppressed.

The phase $\phi$ of the radiation is a statistical quantity; hence, terms in (\ref{A1}) containing
$\alpha^2=\langle\alpha|a^2|\alpha\rangle$ vanish upon averaging on $\phi$ (the same thing happens for $(\alpha^*)^2$,
equally distributed phases on $0,\dots, 2\pi$).
Apart from a very small difference of $|\alpha|^2$ and  $|\alpha|^2+1$, $D^<_{\mu\nu}(\rr,t)$ and $D^>_{\mu\nu} (\rr,t)$
become equal and depend only on $\rr=\rr_1-\rr_2$, $t=t_1-t_2$.
As a result, the retarded and advanced $D$ vanish, and the Keldysh component becomes
\begin{equation}
D^K_{\mu\nu} (\rr,t) = -i \;\frac{I}{\omega^2\epsilon_0 c \eta} \big( \; e_{\mu} \;
   e^*_{\nu} \; e^{i (\qq\rr - \omega t ) } + cc \;\big).
   \label{A2}
\end{equation}
As the light wave length is much larger than the crystal unit cell, we may approximate  $e^{\pm i \qq\rr} \to 1$
(the dipole approximation, neglecting the photon--drag effect). This is result (\ref{Keldysh-D}).
\vfill

\begin{widetext}
\section{Diamagnetic contribution to the tensor $\RR^S$}
In the velocity gauge there is a (small) ``diamagnetic" contribution from the vertex operator
$\frac{q^2}{m_0}\mathbf A_{cl}$ to the generation matrix $G_{nn'}$, which is usually neglected.
In linear order with respect to $\BB$, this contribution reads
\begin{eqnarray}
&\displaystyle\delta G_{n}^{(\BB,dia)}(\kk) = I \frac{\pi q^3}{\omega^3\,m_0^3\,\epsilon_0\,c\,\eta}\,
B_{\beta}\,\epsilon_{\beta\nu\gamma}
\sum_{\stackrel{n'}{\Omega=\pm\omega}}(f_{n',0}(\kk) - f_{n,0}(\kk))\,\,
\delta(E_{n'}(\kk) - E_{n}(\kk) - \Omega)\,\textnormal{sign}(\Omega)\nonumber\\
&\displaystyle\times\big\{\Re\big (\langle n,\kk|p_{\mu}|n',\kk \rangle\,\langle n',\kk|p_{\gamma}|n,\kk \rangle\big )\,
\Im(e^*_{\mu,\Omega}\, e_{\nu,\Omega})+
\displaystyle\Im\big (\langle n,\kk|p_{\mu}|n',\kk \rangle\, \langle n',\kk|p_{\gamma}|n,\kk \rangle\big )\,
\Re(e^*_{\mu,\Omega}\, e_{\nu,\Omega})\big\}.
\label{G-Rate-dia}
\end{eqnarray}
This result is obtained in the same way as $G_{n}^{(0)}$ in Eqs.~(\ref{G-Rate},\ref{Generationsrate}), by
taking into account the terms linear in $\BB$ in the product of the matrix elements
$\langle n,\kk|p_{\mu}-\frac{q}{2}(\BB\times\rr)_{\mu}|n',\kk \rangle\;
 \langle n',\kk|p_{\nu}-\frac{q}{2}(\BB\times\rr)_{\nu}|n,\kk \rangle$
of the vertex operator.
Note that the phase factor contained in the approximation  Eq.~(\ref{G-Approx}) is responsible for
transforming the gauge--dependent field $\mathbf{A}_{cl}$ into the gauge--independent term
$\frac{1}{2}\BB\times\rr$ in the vertex operator.
Subsequently, the matrix element of the position operator $\rr$ is replaced
by that of the momentum operator $\pp$ using the identity
$\langle n,\kk|\rr|m,\kk \rangle = \frac{1}{i\;m_0}
\langle n,\kk|\pp|m,\kk \rangle/(E_n(\kk) - E_m(\kk))$, which holds for $E_n(\kk) \ne E_m(\kk)$.
Moreover,  only odd terms in $\kk$ contribute, i.e., terms containing
$\Im\big(\langle n,\kk\dots n,\kk \rangle\big)\Re(e^*_{\mu,\Omega}\, e_{\nu,\Omega})$,
giving a contribution to $\RR^S$ but not to $\RR^A$.
Following the same route as for  $\PP^A$, cf.~(\ref{df-circular}-\ref{PA}), we obtain
\begin{eqnarray}
&\displaystyle R^{S,dia}_{\alpha\beta\mu\nu} = I \frac{q^4}{4\,\pi^2\,\omega^3\,m_0^3\,\epsilon_0\,c\,\eta}\,
\,\epsilon_{\beta\nu\gamma}
\sum_{\stackrel{n',n}{\Omega=\pm\omega}}\int\limits_{\,1.BZ}\!d^3k\,\,(f_{n',0}(\kk) - f_{n,0}(\kk))\,\,
\delta(E_{n'}(\kk) - E_{n}(\kk) - \Omega)\times\nonumber\\
&\displaystyle\,\tau_n\, v_{n,\alpha}(\kk)\,\big(\delta_{v,n}+\delta_{c,n}\big)\,\,
\Im\big (\langle n,\kk|p_{\mu}|n',\kk \rangle\, \langle n',\kk|p_{\gamma}|n,\kk \rangle\big )\,
\,\sign(\Omega).
\label{tensor-rsdia}
\end{eqnarray}
\end{widetext}
Remarkably, result (\ref{tensor-rsdia}) can be linked to $\PP^A$ by Eq.~(\ref{PA})
\begin{equation}
\displaystyle R^{S,dia}_{\alpha\beta\mu\nu} = \frac{e}{\omega\, m_0}\,
\epsilon_{\beta\nu\gamma}\, P_{\alpha\mu\gamma}^A.
\label{dia-rs}
\end{equation}
Hence, diamagnetic contributions to $\RR^S$ exist only in nongyrotropic media,
yet a different spectral dependence may be expected.

For a crude estimate we consider parabolic valence and conduction bands and disregard the angular dependence
of $\kk$ in Eqs.~(\ref{tensor-rsdia}) and (\ref{RSwxyz}).
Near the energy gap $\Delta$, we have
\[
|\RR^{S,dia}| \approx |\RR^S| (1 - \frac{\Delta}{\omega}), \,\,\omega \ge \Delta.
\]
This result supports the usual approximation to neglect the diamagnetic contribution near the gap.
Nevertheless, it should be taken into account in numerical calculations covering a wide frequency range.

\section{Numerical Application to GaP}

The expressions for the response coefficients,
Eqs.~(\ref{Pxyz},\ref{PA2},\ref{RSwxyz},\ref{RAbal},\ref{RAshift}) involve band energies and momentum matrix
elements which are directly available or can be obtained from band structure calculations.
With respect to the shift mechanism, n-doped GaP is a particularly favorable system.
Optical transitions occur from the bottom of the conduction band (near the X point) to the next upper band
which is separated by a small gap of $\Delta=355$~meV.
The latter is solely due to the noninversion symmetry of  the crystal.
Previous calculations\cite{Hornung-1} for the absorption coefficient and linear photogalvanic tensor component
$P_{xyz}$ proved to be in almost perfect agreement with experimental results.

GaP belongs to the symmetry group $\bar43m$. For $\PP^S$ there is only a single independent element,
$P_{xyz}$, whereas $\PP^A$ vanishes identically because GaP is nongyrotropic.
In this symmetry, a fourth--rank axial tensor has three independent components\cite{Birss},
which are chosen as $R^S_{xxyy}$, $R^S_{xyxy}$, and $R^A_{xyxy}$.

To keep the presentation simple, we use the results for GaP from a local pseudopotential calculation\cite{Hornung-1}.
The conduction band and next upper band near the X point are nondegenerate and there are six pockets with equal occupation.
Band energies are modeled analytically, whereas the $\kk$ dependence of the momentum matrix elements\cite{Hornung-1}
will be neglected.
$\langle c,\kk|p_{\nu}|c^*,\kk\rangle$ ($\nu\hat= x,y$) is solely different from zero in the pockets on the
$\kk_x$ and $\kk_y$ axes.
At room temperature the electron system for $n=2.4\times 10^{16}$cm$^{-3}$ is nondegenerate.

\begin{figure}[h]
\begin{center}
\includegraphics[width=8cm]{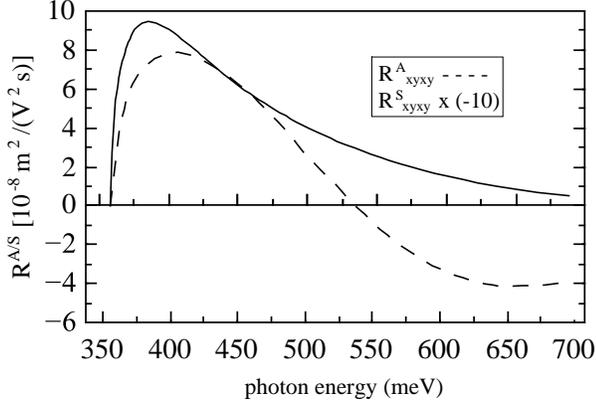}
\caption{Tensor components $R^A_{xyxy}$ and  $R^S_{xyxy}$ \label{gap}}
\end{center}
\end{figure}

Within this approximation (rotationally symmetric energy surfaces) $R^S_{xxyy}$ vanishes, whereas $R^S_{xyxy}$ is nonvanishing,
and a momentum relaxation time of $\tau = 5.0\cdot 10^{-14}s$ has been assumed.

To determine $R^{A,shift}_{xyxy}$, the $\QQ_i$ derivatives ($i=1,2)$ have first to be calculated.
The dominant contribution results from a sum of products whose two factors are first derivatives with respect to $\mathbf{Q_i}$.
One factor contains the Fermi functions and the $\delta$ function,
while the second factor results from  products of matrix elements $ M^{vc/cv}_{xxy}$.
The $\mathbf{Q_i}$ derivatives of the latter are determined using $\kk\cdot\pp$ perturbation theory.
There is no contribution from $\RR^{\textnormal{A,bal}}$ because GaP is nongyrotropic, $R^A_{xyxy} = R^{A,shift}_{xyxy}$.
Numerical results are displayed in Fig.~\ref{gap}. 
The comparatively small numerical values for $R^S_{xyxy}$ and $R^A_{xyxy}$ are due to the low electron concentration $n$.

\vfill

\newpage
\begin{widetext}
\begin{center}
\Large{Supplemental material: Feynman Diagrams} 
\end{center}

We assign the usual graphical symbols to the terms of Eqs.~(\ref{Dyson1}-\ref{G-Approx}):\\

\begin{tabular}{lrcl}
Inverse Green's function&&&\\
including $\AA_{cl}$ and $\Phi_{cl}$:&
 $\mathbf{\hat G}^{-1}_{cl}$\;&=&$\feyn{fA \vertexlabel^{-1}}$
\; = $(i\; \partial_t - H_{cl}(\rr,\pp,t))\cdot\mathbf{\hat 1}$.\\
&&&\\
Green's function with the influence&&&\\
of $\AA_{cl}$, $\Phi_{cl}$ but without $\AA_{rad}$:&
   $\mathbf{\hat G}_{cl}$\;&=& $\feyn{fA}$.\\
&&&\\
Complete Green's function:&
 $\mathbf{\hat G}$\;&=& $\feyn{mA}$.\\
&&&\\
Radiation with vertex (Eq.~(\ref{ham3})):&
$\feyn{xg}$\;\; &=& $-H_{int}(\rr,\pp,t)\cdot\mathbf{\hat 1}$  $=$
$\frac{q}{m_0}(\pp-q\AA_{cl})\cdot \AA_{rad}(t)\cdot\mathbf{\hat 1}$.\\
&&&\\
Dyson equation for $\mathbf{\hat G}$:&
$\feyn{fA \vertexlabel^{-1}\;\; mA}$\;&=&
$\delta(t_1 - t_2) \delta(\rr_1 - \rr_2)\cdot\mathbf{\hat 1}$ +\; $\feyn{xgv mA}$.
\end{tabular}

\noindent\newline%
In the Dyson equation $(\rr,\pp,t)$ stands for $(\rr_1,\pp_1,t_1)$ and $(\rr_2,\pp_2,t_2)$ in its adjoint and
$\mathbf{\hat 1}$ is the unit matrix in Keldysh space.  $\mathbf{\hat G}_{cl}$ plays the role of the "non--interacting" Green's
function (in quasiclassical approximation) with respect to the radiation.
The contribution of the vertex--operator $-q\AA_{cl}$ is treated separately in the Appendix B.

To find the part of $G^K$, which depends only on the "mean" - time T, the Dyson equation is iterated
and only the graphs  with even number of vertices are considered.
Then, by closing the open photon lines in pairs,  graphs with $\Omega’s$ of opposite signs are combined. 
This corresponds to an averaging over time T.

Due to the weak time dependence of the classical fields  $\AA_{cl}$ and $\Phi_{cl}$ these graphs contain
the relevant contributions to the photo--currrent.

As a result, we obtain:

\vspace{12pt}
{\small
 $\feyn{!{mA}{even} =  fA  +  fA !{gv}{\Omega_1} fA !{gv}{\Omega_2} fA +
 fA  !{gv}{\Omega_1} fA !{gv}{\Omega_2} fA  !{gv}{\Omega_3} fA !{gv}{\Omega_4} fA+ \cdots}$

 $\feyn{!{mA}{DC} = fA + fA  f !{glA}\Omega fA fA +  fA  f !{glA}{\Omega_1} fA
  fA  f !{glA}{\Omega_2} fA fA + fA fA !{glA}{\Omega_1} fA !{glA}{\Omega_2} fA fA +
  fA f !{glBA}{\Omega_1}  ![bot]{glSA}{\Omega_2} f fA + \cdots}$
}

\vspace{12pt}
Summation on all $\Omega_i = \pm\omega$ is performed independently.

We are looking for the current--contribution, which is linear in the intensity 
(quadratic in the $\AA_{rad}$--field); therefore, only the first two terms are relevant

\vspace{12pt}
\begin{center}
 $\feyn{fA \vertexlabel^{-1}\;\; !{mA}{DC}} \approx \delta(t) \delta(\rr)\cdot\mathbf{\hat 1} \quad + \quad $
 $\feyn{x f ![bot]{glA}{} fA x fA}$,
\end{center}

and for the same reason  $\mathbf{\hat G}_{cl}$ is approximated by
\begin{equation*}
 \mathbf{\hat G}_{cl}\;\approx\;\mathbf{\hat G_0}(\rr_1,t_1;\rr_2,t_2)
\; e^{iq[(\rr_1-\rr_2)\mathbf A_{cl}(\RR,T)- (t_1-t_2)\Phi_{cl}(\RR,T)]}.
\end{equation*}

\vspace{12pt}
 For the photon line with the attached vertex operators $\pp_{1,\mu}-q\AA_{cl,\mu}(\rr_1)$
  and $\pp_{2,\nu}-q\AA_{cl,\nu}(\rr_2)$ we get:

\vspace{12pt}
\begin{center}
 $\feyn{x !{gA}{} x}$\;\;=  $\frac{q^2}{m_0^2}\;\mathbf{\hat 1}\cdot
  (\pp_{1,\mu}-q\AA_{cl,\mu}(\rr_1))\;\;
  \frac{i}{2} D^K_{\mu\nu}(t_1-t_2)\;\; (\pp_{2,\nu}-q\AA_{cl,\nu}(\rr_2)) \cdot  \mathbf{\hat 1}$.
\end{center}

For $D^K$ see Eq.~(\ref{Keldysh-D}).
The self--energy in this approximation is:

\vspace{12pt}
\begin{center}
\begin{tabular}{rcl}
&$\feyn{x f glA fA x}\;\;\approx\;\;\frac{q^2}{m_0^2}\;\frac{i}{2}\; D^K_{\mu\nu}(t_1-t_2)$
$\mathbf{\hat 1}\cdot  (\pp_{1,\mu}-q\AA_{cl,\mu}(\rr_1))$&\\
\tabularnewline[-0.3cm]
 &$\times\{\mathbf{\hat G_0}(\rr_1,t_1;\rr_2,t_2)
\; e^{iq[(\rr_1-\rr_2)\mathbf A_{cl}(\RR,T)- (t_1-t_2)\Phi_{cl}(\RR,T)]}\} (\pp_{2,\nu}-q\AA_{cl,\nu}(\rr_2))
\cdot  \mathbf{\hat 1}
.$&
\end{tabular}
\end{center}

\vspace{12pt}
The Keldysh rules for Feynman diagrams  are more complicated than expected, see Rammer and Smith (Ref.\cite{Rammer}, 
their Eqs.~(2.39-2.43)). Due to the special structure of the photon Keldysh matrix (only $D_{12}$ is nonzero) 
there are two unit matrices on the vertices and a factor $1/2$.

To identify the Hermitian and anti--Hermitian parts  of the self--energy, we use Eqs.~(\ref{G0}) and (\ref{g0}).
Equation~(\ref{g0}) is fouriertransformed with respect to the relative time $t$. 
Then the result is decomposed into real and imaginary parts, which are directly related to
the Hermitian and anti--Hermitian parts of the self--energy.
\end{widetext}

\begin{thebibliography}{99}
\bibitem{comment-1} In Ferroelectrics, circular polarized light also contributes via $\PP^S$, like unpolarized light.
                    Ballistic currents belong to the broader class of "injection" currents\cite{Sipe-1},
                    which would rise linearly with time in the absense of damping.
\bibitem{twobands} A two band situation is implied. In general, all pairs of direct optical transition have to be
			      taken into account additively.
\bibitem{typical} Results are often given in terms of a second--order conductivity
                  $j_{\alpha}=\sigma_{\alpha\mu\nu}\,\bar{E}_\mu\,\bar{E}_\nu$,
                  where $\bar{E}_{\mu}$ are the components of the electrical field amplitude of the light in the material.
                  Conversion is $I = c\epsilon_0\,\bar{E}^2\eta/2$, $\bar{E}^2 \approx 750 (V/A)\,I/\eta$,
                  where $\eta$ is the refractive index.
                  Best Si solar cells have an effective $\sigma_{eff}\sim1330\eta\,\mu$A/V$^2$, ($P_{eff}\sim1/V$).
                  From Ref.~\cite{Rangel}:
                  BaTiO$_3$ (exp.): $\sigma_{zzz}\approx 5\mu$A/V$^2$, ($P^S_{zzz}\approx4\times10^{-3}$/V);
                  GeS (theor.): $\sigma_{zxx}\approx100\mu$A/V$^2$, ($P^S_{zxx}\approx0.075$/V).
\bibitem{spurious} Improper treatment of these terms may lead to spurious effects, as e.g.
        in Ref.\cite{Kristoffel-2} or mentioned by Refs.\cite{Sipe-1,Ivchenko-4}.
\bibitem{comment-2} "Shift" and "ballistic" refer to different generation processes. The corresponding
         currents don't have different electro--mechanical properties.
\end{thebibliography}
\end{document}